\documentstyle[12pt,psfig]{article}
\begin{document}

\newcommand{\be}{\begin{equation}}
\newcommand{\ee}{\end{equation}}
\newcommand{\bea}{\begin{eqnarray}}
\newcommand{\eea}{\end{eqnarray}}
\newcommand{\gt}{\widetilde{g}}
\newcommand{\Tt}{\widetilde{T}}
\newcommand{\lag}{{\cal L}}
\newcommand{\rhot}{\widetilde{\rho}}
\newcommand{\pt}{\widetilde{p}}
\newcommand{\rhodyn}{\rho_{\rm dyn}}
\newcommand{\pdyn}{p_{\rm dyn}}
\newcommand{\4}{{(4)}}
\newcommand{\5}{{(5)}}
\newcommand{\M}{{\rm M}}
\newcommand{\R}{{\rm R}}
\newcommand{\vac}{{\rm vac}}
\newcommand{\dark}{{\rm dark}}
\newcommand{\total}{{\rm total}}
\newcommand{\planck}{{\rm Planck}}

\baselineskip 16pt

\begin{flushright}
astro-ph/0107571 \\ EFI-2001-27  \end{flushright} 

\vspace*{2cm}

\begin{center}

{\Large{\bf Dark Energy and the Preposterous Universe}}

\vspace*{0.3in}
Sean M. Carroll
\vspace*{0.3in}

\it
Enrico Fermi Institute and Department of Physics,
University of Chicago\\
5640 S.~Ellis Avenue, Chicago, IL~60637, USA\\
{\tt carroll@theory.uchicago.edu} \\
\vspace*{0.2in}

\end{center}

\begin{abstract}
A brief review is offered
of the theoretical background concerning dark
energy: what is required by observations, what sort of models are
being considered, and how they fit into particle physics and
gravitation.  Contribution to the SNAP (SuperNova Acceleration Probe) 
Yellow Book.
\end{abstract}

\vfill

\newpage

\section{The Preposterous Universe}

Surprising experimental results are the most common driving force
behind significant advances in scientific understanding.  The recent
discovery that the universe appears to be dominated by a component of
``dark energy'' qualifies as an extraordinarily surprising result; we
have every reason to be optimistic that attempts to understand this
phenomenon will lead to profound improvements in our pictures of
gravitation, particle physics, and gravitation.

\subsection{Dark energy}

In general relativity,
a homogeneous and isotropic universe is characterized by two 
quantities, the spatial curvature $\kappa$ and scale factor $a(t)$.
These are related to the energy density $\rho$ by the Friedmann
equation:
\be
  H^2 \equiv \left({\dot a \over a}\right)^2 =
  {8\pi G\over 3}\rho - {\kappa \over a^2}\ .
  \label{feq1}
\ee
For any value of the Hubble expansion parameter $H$, there is
a critical density which solves this equation for zero spatial
curvature: $\rho_{\rm crit} = 3H^2/8\pi G$.  The energy density
is conveniently characterized by a density parameter constructed
by normalizing with respect to the critical density:
$\Omega = \rho/\rho_{\rm crit}$.  

Observations of the dynamics
of galaxies and clusters have shown that the amount of ``matter''
(slowly-moving particles that can fall into local gravitational
potential wells) is $\Omega_\M = 0.3 \pm 0.1$, short of the
critical density.  At the same time, however, observations of
temperature anisotropies in the cosmic microwave background (CMB)
are consistent with nearly scale-free, gaussian, adiabatic
primordial density perturbations (the kind predicted by the
inflationary universe scenario) for a nearly spatially flat
universe, $\Omega_\total \approx 1$.  We therefore infer the
existence of a dark energy component $\rho_\dark$ 
smoothly distributed
through space (so that it does not influence the local motions
of galaxies and clusters), with $\Omega_\dark \approx 0.7$.
(See \cite{carroll} for a recent overview and references.)

Meanwhile, measurements of the distance
vs.\ redshift relation for Type Ia supernovae \cite{riess,perl}
have provided evidence that the universe is accelerating ---
that $\ddot a > 0$.  The significance of this discovery can be
appreciated by rewriting the Friedmann equation (\ref{feq1})
after multiplying by $a^2$:
\be
  {\dot a}^2 =
  {8\pi G\over 3}a^2\rho - \kappa\ .
  \label{feq2}
\ee
The energy density in matter (non-relativistic particles)
diminishes as the number density is diluted by expansion, so that
$\rho_\M \propto a^{-3}$.  If particles are relativistic, and
thus classified as ``radiation'', they are both diluted in number
density and have their individual energies redshift as $a^{-1}$, 
so that $\rho_\R \propto a^{-4}$.  For either of these conventional
sources of energy density, the right-hand side of 
(\ref{feq2}) will be decreasing in an expanding universe (since
$a^2 \rho$ is decreasing, while $\kappa$ is a constant), 
so that $\dot a$ will be decreasing.
The supernova data therefore imply that, to make the
universe accelerate, the dark energy must be 
varying slowly with time (roughly speaking,
redshifting away more slowly than $a^{-2}$) as well as with space.  

There is a straightforward candidate for a dark energy component
that varies slowly in both space and time: vacuum energy, or
the cosmological constant (for reviews see
\cite{carroll,weinberg,cpt,cohn,sahni,witten}).  The distinguishing
feature of vacuum energy is that it is a minimum amount of energy
density in any region, strictly constant throughout spacetime.
To match the data, we require a vacuum energy
\be
  \rho_\vac \approx 
  (10^{-3}{\rm ~eV})^4 = 10^{-8} {\rm ~ergs/cm}^3\ .
  \label{rhoobs}
\ee
(In units where $\hbar = c =1$, energy density has units
of [energy]$^4$.)
The idea that the dark energy density is simply a constant inherent
in the fabric of spacetime is in excellent agreement with the data,
but raises two very difficult questions: first, why is the vacuum
energy so much smaller than what we would think of as its natural
value (the cosmological constant problem); 
and second, why are the matter and vacuum energy densities
approximately equal today (the coincidence problem)?  
Of course the first question is 
important even if the dark energy is not a cosmological constant,
although a nonzero value for the vacuum energy makes its smallness
perhaps even more puzzling than if it were simply zero.

\subsection{The cosmological constant problem}

Let us turn first to the issue of why the vacuum energy is
smaller than we might expect.  Although the notion that empty space
has a nonzero energy density can seem surprising at first, it is
a very natural occurrence in any generic pairing of general relativity
with field theory (quantum or classical).  
We can consider for definiteness a simple model
of a single real scalar field $\phi$ with a potential energy density
$V(\phi)$.  The total energy density is
\be
  \rho_\phi = {1\over 2}{\dot \phi}^2 + {1\over 2}(\nabla\phi)^2
  + V(\phi)\ ,
\ee
where $\nabla$ represents the spatial gradient.
It is immediately clear that any solution in which the field takes
on a constant value $\phi_0$
throughout spacetime will have an energy density which is constant
throughout spacetime, $\rho_\phi = V(\phi_0)$.  The crucial point
is that there is no principle or symmetry in such a theory
which would prefer that $V(\phi_0)$ have the value zero rather than
any other value.  In richer theories there may be such principles,
such as supersymmetry or conformal invariance; the observed world,
however, shows no sign of such symmetries, so they must be
severely broken if they exist at all.  Hence, it requires fine-tuning
to obtain a vanishing $\rho_\vac$.

We are unable to reliably calculate the expected vacuum energy in
the real world, or even in some specific field theory such as the 
Standard Model of particle physics; at best we can offer 
order-of-magnitude estimates for the contributions from different
sectors.  In the Standard Model there are at least two important
contributions, from nonvanishing condensates in the vacuum: the
potential energy of the Higgs field, expected to be of the order
$(100 {\rm ~GeV})^4 = (10^{11} {\rm ~eV})^4$, 
and a QCD energy density in the condensate
of quark bilinears $\bar{q}q$ responsible for chiral symmetry breaking,
expected to be of the order $(100 {\rm ~MeV})^4 = (10^{8} {\rm ~eV})^4$.
There is also a contribution from the quantum-mechanical
zero-point vacuum fluctuations of each field
in the model.  This contribution actually diverges due to effects
of very high-frequency modes; it is necessary to introduce a cutoff and
hope that a more complete theory will eventually provide a physical 
justification for doing so.  If this cutoff is at the Planck scale
$M_\planck = {1/\sqrt{8\pi G}} = 10^{18}{\rm ~GeV}$, we obtain a
vacuum energy of order $(10^{18} {\rm ~GeV})^4 = (10^{27} {\rm ~eV})^4$.
Similarly, there is no reason to exclude a ``bare'' classical 
contribution to the cosmological constant at the Planck scale,
$\rho_{\Lambda_0} \sim (10^{18} {\rm ~GeV})^4$.  For any of these
examples, we cannot even say with confidence whether the 
corresponding energy density is positive or negative; nevertheless,
since there is no apparent relationship between the values of the
disparate contributions, we expect the total vacuum energy
to be of the same order as that of the largest components:
\be
  \rho_\vac^{{\rm (theory)}} \sim (10^{27} {\rm ~eV})^4
  = 10^{112}{\rm ~ergs/cm}^3\ .
  \label{rhotheory}
\ee

There is clearly a mismatch between the theoretical prediction
(\ref{rhotheory}) and the observed value (\ref{rhoobs}):
\be
  \rho_\vac^{{\rm (theory)}} \sim 10^{120} \rho_\vac^{{\rm (obs)}}\ .
\ee
This is the famous 120-orders-of-magnitude discrepancy that makes
the cosmological constant problem such a glaring embarrassment.
Of course, it is somewhat unfair to emphasize the factor of
$10^{120}$, which depends on the fact that energy density
has units of [energy]$^4$.  If we express the vacuum energy in
terms of a mass scale, $\rho_\vac = M_\vac^4$, the discrepancy becomes
$M_\vac^{{\rm (theory)}} \sim 10^{30} M_\vac^{{\rm (obs)}}$; it is 
more accurate to think of the cosmological constant problem as a
discrepancy of 30 orders of magnitude in energy scale.  In fact,
this problem can be ameliorated in theories where 
supersymmetry is spontaneously broken at a low scale, since the
vacuum energy will then be given by the scale at which supersymmetry
is broken (above that energy, for example, the zero-point
contributions from fermions are exactly canceled by equal and opposite
contributions from bosonic superpartners).  If supersymmetry
is preserved down to just above the weak scale, so that 
$M_\vac \approx M_{\rm SUSY} \approx 10^3{\rm ~GeV}$, we would have
$M_\vac^{({\rm SUSY})} = 10^{15}M_\vac^{\rm (obs)}$.  In the most
optimistic reading, therefore, we are left with a discrepancy of
a mere fifteen orders of magnitude that we have no idea how to
resolve; still, this qualifies as
a problem worthy of our attention.

There have been a large number of suggested resolutions to the
cosmological constant problem; see 
\cite{carroll,weinberg,cohn,sahni,witten}
for reviews.  To date none has seemed exceptionally compelling, and
most researchers believe that the correct solution has yet to be
found.

\subsection{The coincidence problem}

The second issue mentioned above is the coincidence between the
observed vacuum energy (\ref{rhoobs}) and the current matter
density.  The ``best-fit universe'' model has 
$\Omega_{\Lambda 0} = 0.7$ and $\Omega_{\M 0} = 0.3$, but the
relative balance of vacuum and matter changes rapidly as the
universe expands:
\be
  {\Omega_\Lambda \over \Omega_\M} = {\rho_\Lambda
  \over \rho_\M} \propto a^3\ .
\ee
As a consequence,
\begin{figure}[t]
  \centerline{
  \psfig{figure=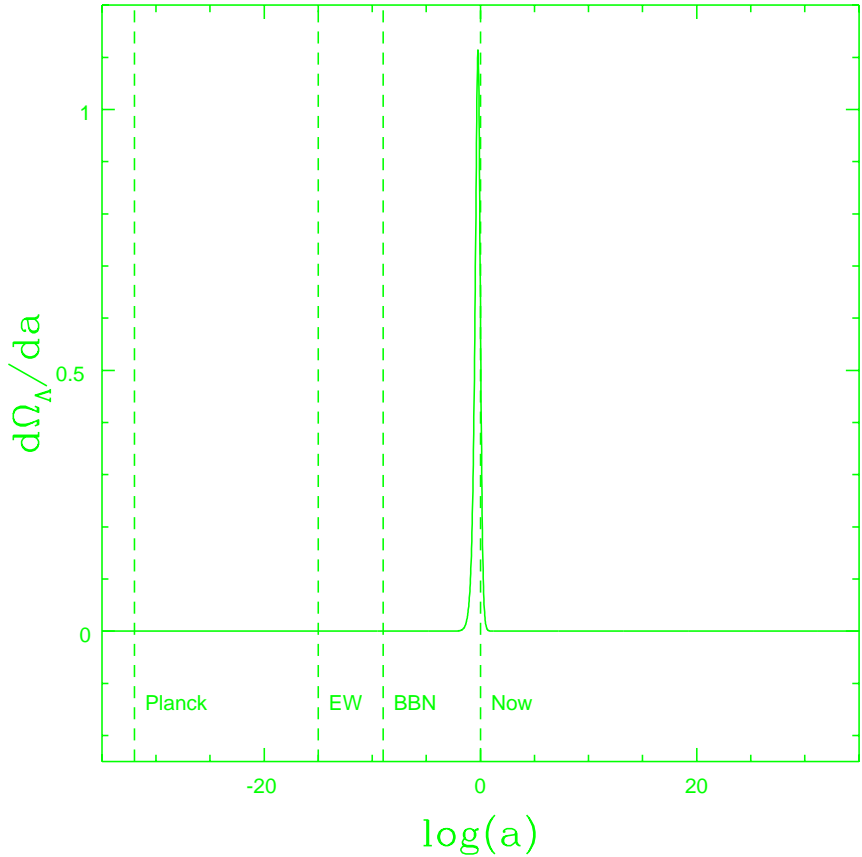,height=12cm}}
  \caption{The rate of change of the vacuum energy density parameter,
  $d\Omega_\Lambda/da$, as a function of the scale factor $a$, in
  a universe with $\Omega_{\Lambda 0} = 0.7$, $\Omega_{\M 0} = 0.3$.
  Scale factors corresponding to the Planck era, electroweak symmetry
  breaking (EW), and Big Bang nucleosynthesis (BBN) are indicated, as
  well as the present day.
  The spike reflects the fact that, in such a universe, there is only
  a short period in which $\Omega_\Lambda$ is evolving noticeably
  with time.}
  \label{doldavsa2}
\end{figure}
at early times the vacuum energy was negligible in comparison to
matter and radiation, while at late times matter and radiation are
negligible.  There is only a brief epoch of the universe's history
during which it would be possible to
witness the transition from domination by
one type of component to another.  This is illustrated in 
Figure~\ref{doldavsa2}, in which the rate of change of
$\Omega_\Lambda$ is plotted as a function of the scale factor.
At early times $\Omega_\Lambda$ is close to zero and changing
very slowly, while at late times it is close to unity and 
changing very slowly.  It seems remarkable that we live during
the short transitional period between these two eras.

The approximate coincidence between matter and vacuum energies in the
current universe is one of several puzzling features of the
composition of the total energy density.  Another great surprise is
the comparable magnitudes of the baryon density ($\Omega_{\rm b}
\approx 0.04$) and the density of cold non-baryonic dark matter
($\Omega_{\rm CDM}\approx 0.25$), and perhaps also that in massive
neutrinos ($\Omega_\nu \leq 0.01$).  In our current understanding,
these components are relics of completely unrelated processes in the
very early universe, and there seems to be no good reason why they
should be of the same order of magnitude (although some specific
models have been proposed).  The real world seems to be a more rich
and complex place than Occam's razor might have predicted.  It is
important to keep in mind, however, the crucial distinction between
the coincidences relating the various matter components and that
relating the matter and vacuum energy: the former are set once and for
all by primordial processes and remain unchanged as the universe
evolves, while the latter holds true only during a certain era.  It is
fruitless to try to explain the matter/vacuum coincidence by invoking
mechanisms which make the dark energy density time-dependent in such a
way as to {\it always} be proportional to that in matter; such a
scenario would either imply that the dark energy would redshift away
as $\rho_\dark \propto a^{-3}$, which from (\ref{feq2}) would lead to
a non-accelerating universe, or require dramatic departures from
conventional general relativity, which would in turn make it difficult
to recover the successes of conventional cosmology (Big Bang
nucleosynthesis, CMB anisotropy, growth of structure, and the age of
the universe, to name a few).  Recent observations provide some
evidence that the universe has only recently entered an era of
acceleration out of a previous era of deceleration \cite{97ff};
although the observational case is not airtight, the conclusion seems
inescapable.

\section{What might be going on?}

It may seem misguided to put a great deal of energy into exploring
models of a small nonzero dark energy density when we have very
little idea why the vacuum energy is not as large as the Planck
scale.  On the other hand, the discovery of dark energy may provide
an invaluable clue in our attempts to solve this long-lasting
puzzle, giving us reason to redouble our efforts.  Explanations of
the current acceleration of the universe can be categorized into
one of three types:
\begin{enumerate}
\item The dark energy is a true cosmological constant, strictly
  unchanging throughout space and time.  The minimum-energy
  configuration of the universe may have a small but nonvanishing
  energy density, or we may live in a false vacuum,
  almost degenerate with the true one but with a small nonzero
  additional energy.
\item The cosmological constant is zero, but a slowly-varying 
  dynamical component is mimicking a nonzero vacuum energy.
\item Einstein was wrong, and the Friedmann equation does not
  describe the expansion of the universe.
\end{enumerate}
We briefly examine each of these possibilities in turn.

\subsection{An honest cosmological constant}

The simplest interpretation of the dark energy is that we have
discovered that the cosmological constant is not quite zero: we
are in the lowest energy state possible (or, more properly, that
the particles we observe are excitations of such a state)
but that energy does not vanish.  Although simple, this scenario
is perhaps the hardest to analyze without an understanding of
the complete cosmological constant problem, and there is
correspondingly little to say about such a possibility.  As targets
to shoot for, various numerological coincidences have been 
pointed out, which may some day find homes as predictions of an
actual theory.  For example, the observed vacuum energy scale
$M_\vac = 10^{-3}$~eV
is related to the 1~TeV scale of low-energy supersymmetry breaking
models by a ``supergravity suppression factor'':
\be
  M_\vac = \left({M_{\rm SUSY} \over M_\planck}\right)M_{\rm SUSY}\ .
\ee
In other words, $M_{\rm SUSY}$ is the geometric mean of $M_\vac$
and $M_\planck$.  Unfortunately, nobody knows why this should be
the case.  In a similar spirit, the vacuum energy density is
related to the Planck energy density by the kind of suppression
factor familiar from instanton calculations in gauge theories:
\be
  M_\vac^4 = e^{-2/ \alpha} M_\planck^4\ .
\ee
In other words, the natural log of $10^{120}$ is twice 137.
Again, this is not a relation we have any right to expect to
hold (although it has been suggested that nonperturbative
effects in non-supersymmetric string theories could lead to such
an answer \cite{harvey}).

Theorists attempting to build models of a small nonzero vacuum
energy must keep in mind the requirement of remaining compatible
with some as-yet-undiscovered solution to the cosmological
constant problem.  In particular, it is certainly insufficient
to describe a specific contribution to the vacuum energy which 
by itself is of the right magnitude; it is necessary at the same
time for there to be some plausible reason why the well-known
and large contributions from the Standard Model could be suppressed,
while the new contribution is not.
One way to avoid this problem is to imagine that an unknown
mechanism sets the vacuum energy to zero in the state of lowest
energy, but that we actually live
in a distinct false vacuum state, almost but not quite degenerate
in energy with the true vacuum \cite{tw,gc,kl}.  From an observational
point of view, false vacuum energy 
and true vacuum energy are utterly indistinguishable --- they both appear
as a strictly constant dark energy density.  The issue with such
models is why the splitting in energies between the true and
false vacua should be so much smaller than all of the characteristic
scales of the problem; model-building approaches generally invoke
symmetries to suppress some but not all of the effects that could
split these levels.

The only theory (if one can call it that) which leads a vacuum
energy density of approximately the right order of magnitude
without suspicious fine-tuning is the anthropic principle ---
the notion that intelligent observers will not witness the full
range of conditions in the universe, but only those conditions which
are compatible with the existence of such observers.  Thus, we do
not consider it unnatural that human beings evolved on the surface of
the Earth rather than on that of the Sun, even though the surface
area of the Sun is much larger, since the conditions are rather
less hospitable there.  If, then, there exist distinct parts of
the universe (whether they be separate spatial regions or branches
of a quantum wavefunction) in which the vacuum energy takes on
different values, we would expect to observe a value which favored the 
appearance of life.  Although most humans don't think of the
vacuum energy as playing any role in their lives, a substantially
larger value than we presently observe would either have led to
a rapid recollapse of the universe (if $\rho_\vac$ were negative)
or an inability to form galaxies (if $\rho_\vac$ were positive).
Depending on the distribution of possible values of $\rho_\vac$,
one can argue that the recently observed value is in excellent
agreement with what we should expect \cite{vilenkin,efstathiou,msw,gv}.
Many physicists find it unappealing to think that an apparent
constant of nature would turn out to simply be a feature of our
local environment that was chosen from an ensemble of possibilities,
although we should perhaps not expect that the universe takes our
feelings into account on these matters.
More importantly, relying on the anthropic principle involves
the invocation of a large collection of alternative possibilities
for the vacuum energy, closely spaced in energy but not continuously 
connected to each other (unless the light scalar
fields implied by such connected vacua is very weakly
coupled, as it must also be in the quintessence models discussed
below).  It is by no means an
economical solution to the vacuum energy puzzle.

As an interesting sidelight to this issue, it has been claimed that
a positive vacuum energy would be incompatible with our current
understanding of string theory \cite{bf,hks,fkmp,bd}.  At issue is
the fact that such a universe eventually approaches a de~Sitter
solution (exponentially expanding), which implies future horizons
which make it impossible to derive a gauge-invariant S-matrix.
One possible resolution might involve a dynamical 
dark energy component such as
those discussed in the next section.  While few string theorists
would be willing to concede that a definitive measurement that the
vacuum energy is constant with time would rule out string theory as
a description of nature, the possibility of saying something important
about fundamental theory from cosmological observations presents an
extremely exciting opportunity.

\subsection{Dynamical dark energy}

Although the observational evidence for dark energy implies a 
component which is unclustered in space as well as slowly-varying
in time, we may still imagine that it is not perfectly
constant.  The simplest possibility along these lines
involves the same kind of source
typically invoked in models of inflation in the very early universe:
a scalar field rolling slowly in a potential, sometimes known as
``quintessence'' \cite{pr,pngb,q}.  There are also a number of more
exotic possibilities, including tangled topological defects and
variable-mass particles (see \cite{carroll,sahni} for references and
discussion).

There are good reasons to consider dynamical dark
energy as an alternative to an honest cosmological constant.
First, a dynamical energy density can be evolving slowly to zero,
allowing for a solution to the cosmological constant problem which
makes the ultimate vacuum energy vanish exactly.  Second, it poses
an interesting and challenging observational problem to study the
evolution of the dark energy, from which we might learn something
about the underlying physical mechanism.  Perhaps most intriguingly,
allowing the dark energy to evolve opens the possibility
of finding a dynamical solution to the coincidence problem, if the
dynamics are such as to trigger a recent takeover by the dark energy
(independently of, or at least for a wide range of, the 
parameters in the theory).

At the same time, introducing dynamics opens up the possibility
of introducing new problems, the form and severity
of which will depend on the specific
kind of model being considered.  The most popular quintessence
models feature scalar fields $\phi$ with masses of order the 
current Hubble scale,
\be
  m_\phi \sim H_0 \sim 10^{-33} {\rm ~eV}\ .
\ee
(Fields with larger masses would typically have already rolled
to the minimum of their potentials.)
In quantum field theory, light scalar fields are
unnatural; renormalization effects tend to drive scalar masses
up to the scale of new physics.  The well-known hierarchy
problem of particle physics amounts to asking why the Higgs
mass, thought to be of order $10^{11}$~eV, should be so much
smaller than the grand unification/Planck scale, 
$10^{25}$-$10^{27}$~eV.  Masses of $10^{-33}$~eV are 
correspondingly harder to understand.  At the same time, such a
low mass implies that $\phi$ gives rise to a long-range force;
even if $\phi$ interacts with ordinary matter only through
indirect gravitational-strength couplings, searches
for fifth forces and time-dependence of coupling constants 
should have already enabled us to detect the quintessence
field \cite{world}.

The need for delicate fine-tunings of masses and couplings in 
quintessence models is certainly a strike against them, but
is not a sufficiently serious one that the idea is not worth
pursuing; until we understand much more about the dark energy,
it would be premature to rule out any idea on the basis of
simple naturalness arguments.  One promising route to gaining
more understanding is to observationally characterize the time
evolution of the dark energy density.  In principle any 
behavior is possible, but it is sensible to choose a simple
parameterization which would characterize
dark energy evolution in the measurable regime of relatively
nearby redshifts (order unity or less).  For this purpose it is
common to imagine that the dark energy evolves as a power law
with the scale factor:
\be
  \rho_\dark \propto a^{-n} \ .
  \label{neq}
\ee
Even if $\rho_\dark$ is not strictly a power law, this ansatz
can be a useful characterization of its effective behavior at
low redshifts.  It is common to define an equation-of-state
parameter relating the energy density to the pressure,
\be
  p = w \rho\ .
  \label{eos}
\ee
Using the equation of energy-momentum
conservation,
\be
  \dot\rho = -3(\rho + p){{\dot a}\over a}\ ,
\ee
a constant exponent $n$ of (\ref{neq}) implies a constant $w$ with
\be
  n = 3(1+w)\ .
\ee
As $n$ varies from 3 (matter) to 0 (cosmological constant), $w$
varies from $0$ to $-1$.  (Imposing mild energy conditions implies
that $|w|\leq 1$ \cite{garnavich}; however, models with $w<-1$ are
still worth considering \cite{caldwell}.)
\begin{figure}[t]
  \centerline{
  \psfig{figure=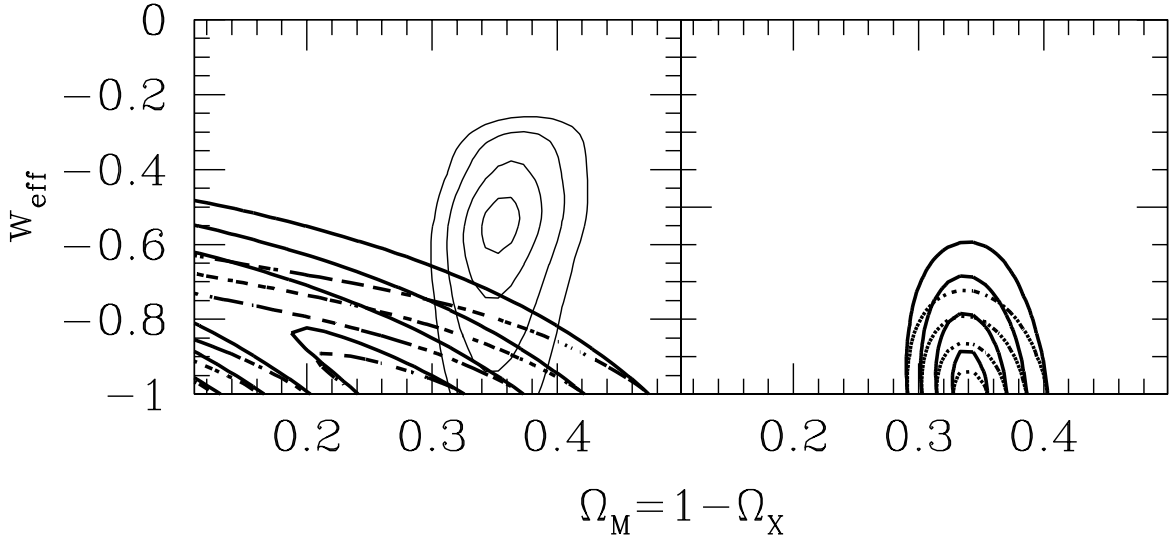,height=8cm}}
  \caption{Limits on the equation-of-state parameter $w$ in a flat
  universe, where $\Omega_\M + \Omega_X = 1$.  The left-hand panel
  shows limits from supernova data (lower left corner) and large-scale
  structure (ellipses); the right-hand panel shows combined 
  constraints.  From \cite{ptw}.}
  \label{wlimits}
\end{figure}
Some limits from supernovae and large-scale structure from \cite{ptw}
are shown in Figure (\ref{wlimits}).  These constraints
apply to the $\Omega_\M$-$w$ plane, under the assumption that the
universe is flat ($\Omega_\M + \Omega_\dark = 1$).  We see that
the observationally favored region features $\Omega_\M \approx
0.35$ and an honest cosmological constant, $w=-1$.  However, there
is plenty of room for alternatives; one of the most important tasks
of observational cosmology will be to reduce the error regions on plots
such of these to pin down precise values of these parameters.

To date, many investigations have considered scalar fields with
potentials that asymptote gradually to zero, of the form
$e^{1/\phi}$ or $1/\phi$.  These can have cosmologically interesting
properties, including ``tracking'' behavior that makes the current
energy density largely independent of the initial conditions
\cite{zlatev}; they
can also be derived from particle-physics models, such as
the dilaton or moduli of string theory.  They do not, however,
provide a solution to the coincidence problem, as the era in which
the scalar field begins to dominate is still set by finely-tuned
parameters in the theory.  There have been
two scalar-field models which come closer to being solutions: 
``$k$-essence'', and oscillating dark energy.  The $k$-essence idea 
\cite{armen} does not put the field in a shallow potential, but
rather modifies the form of the kinetic energy.  We imagine that
the Lagrange density is of the form
\be
  {\cal L} = f(\phi) g(X)\ ,
\ee
where $X={1\over 2}(\nabla_\mu\phi)^2$ is the conventional 
kinetic term.  For certain choices of the functions $f(\phi)$ and
$g(X)$, the $k$-essence field naturally tracks the evolution of
the total radiation energy density during radiation domination,
but switches to being almost constant once matter begins to
dominate.  In such a model the coincidence problem is explained
by the fact that matter/radiation equality was a relatively 
recent occurrence (at least on a logarithmic scale).  The oscillating
models \cite{dks} involve ordinary kinetic terms and potentials,
but the potentials take the form of a decaying exponential with
small perturbations superimposed:
\be
  V(\phi) = e^{-\phi}[1 + \alpha\cos(\phi)]\ .
\ee
On average, the dark energy in such a model will track that of
the dominant matter/radiation component; however, there will be
gradual oscillations from a negligible density to a dominant
density and back, on a timescale set by the Hubble parameter.
Consequently, in such models the 
acceleration of the universe is just something that
happens from time to time.  Unfortunately, in neither
the $k$-essence models nor the oscillating models do we have a
compelling particle-physics motivation for the chosen dynamics,
and in both cases the behavior still depends sensitively on the
precise form of parameters and interactions chosen.  Nevertheless,
these theories stand as interesting attempts to address the 
coincidence problem by dynamical means.

Rather than constructing models on the basis of cosmologically
interesting dynamical properties, we may take the complementary
route of considering which models would appear most sensible from
a particle-physics point of view, and then exploring what 
cosmological properties they exhibit.  An acceptable particle
physics model of quintessence would be one in which the scalar
mass was naturally small and its coupling to ordinary matter
was naturally suppressed.  These requirements are met by
Pseudo-Nambu-Goldstone bosons (PNGB's) \cite{pngb}, which arise in 
models with approximate global symmetries of the form
\be
  \phi \rightarrow \phi + {\rm constant}.
\ee
Clearly such a symmetry should not be exact, or the potential would
be precisely flat; however, even an approximate symmetry can
naturally suppress masses and couplings.  PNGB's typically
arise as the angular degrees of freedom in Mexican-hat
potentials that are ``tilted'' by a small explicitly symmetry
breaking, and the PNGB potential takes on a sinusoidal form:
\be
  V(\phi) = \mu^4[1+ \cos(\phi)]\ .
\ee
As a consequence, there is no easily characterized tracking or
attractor behavior; the equation of state parameter $w$ will
depend on both the potential and the initial conditions, and
can take on any value from $-1$ to $0$ (and in fact will change
with time).  We therefore find that
the properties of models which are constructed by taking
particle-physics requirements as our primary concern appear
quite different from those motivated by cosmology alone.  The
lesson to observational cosmologists is that a wide variety of
possible behaviors should be taken seriously, with data providing
the ultimate guidance.

\subsection{Was Einstein wrong?}

\begin{figure}[t]
  \centerline{
  \psfig{figure=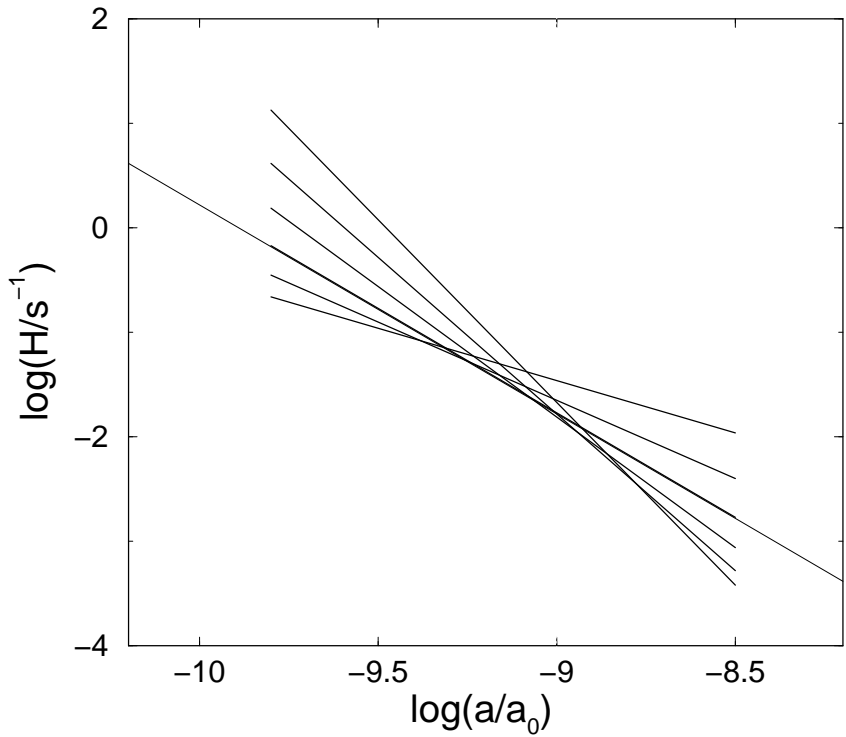,height=10cm}}
  \caption{The range of allowed evolution histories during Big Bang
  nucleosynthesis (between temperatures of 1~MeV to 50~keV),
  expressed as the behavior of the Hubble parameter
  $H=\dot{a}/a$ as a function of $a$.  Changes in the normalization
  of $H$ can be compensated by a change in the slope while 
  predicting the same abundances of $^4$He, $^2$D, and $^7$Li.
  The extended thin line represents the standard radiation-dominated
  Friedmann universe model.  From \cite{ck}.}
  \label{bbnfig}
\end{figure}

Given the uncomfortable tension between observational evidence for
dark energy on one hand and our intuition for what seems 
natural in the context of the standard cosmological model 
on the other, there is an irresistible 
temptation to contemplate the possibility that we are witnessing
a breakdown of the Friedmann equation of conventional 
general relativity (GR) rather than merely a novel
source of energy.  Alternatives to GR are highly
constrained by tests in the solar system and in binary pulsars;
however, if we are contemplating the space of all conceivable
alternatives rather than examining one specific proposal, we are
free to imagine theories which deviate on cosmological scales while
being indistinguishable from GR in small stellar systems.
Speculations along these lines are also constrained by
observations: any alternative must predict the right abundances
of light elements from Big Bang nucleosynthesis (BBN), the correct 
evolution of a sensible spectrum of primordial density fluctuations
into the observed spectrum of temperature anisotropies in the 
Cosmic Microwave Background and the power spectrum of large-scale
structure, and that the age of the universe is approximately twelve
billion years.  Of these phenomena, the sharpest test of 
Friedmann behavior comes from BBN, since perturbation growth 
depends both on the scale factor and on the local gravitational
interactions of the perturbations, while a large number of
alternative expansion histories could in principle give the same
age of the universe.  As an example, Figure (\ref{bbnfig})
provides a graphical representation of alternative expansion 
histories in the vicinity of BBN ($H_{\rm BBN} \sim 0.1 {\rm ~sec}^{-1}$)
which predict the same light element abundances as the standard
picture \cite{ck}.  The point of this figure is that expansion 
histories which are not among the family portrayed, due to 
differences either in the slope or the overall normalization, 
will not give the right abundances.  So it is possible to find
interesting nonstandard cosmologies which are consistent with
the data, but they describe a small set in the
space of all such alternatives.

Rather than imagining that gravity follows the predictions of
standard GR in localized systems but deviates in cosmology, another
approach would be to imagine that GR breaks down whenever the
gravitational field becomes (in some sense) sufficiently weak.
This would be unusual behavior, as we are used to thinking of
effective field theories as breaking down at high energies and
small length scales, but being completely reliable in the 
opposite regime.  On the other hand, we might be ambitious enough to
hope that an alternative theory of gravity could explain away 
not only the need for dark energy but also that for dark matter.
It has been famously pointed out by Milgrom \cite{milgrom}
that the observed dynamics of galaxies only requires the 
introduction of dark matter in regimes where the acceleration
due to gravity (in the Newtonian sense) falls below a certain
fixed value,
\be
  a/c \leq 10^{-18} {\rm ~sec}^{-1}\ .
\ee
Meanwhile, we seem to need to invoke dark energy when the Hubble
parameter drops approximately to its current value,
\be
  H_0 \approx  10^{-18} {\rm ~sec}^{-1}\ .
\ee
{\it A priori}, there seems to be little reason to expect that
these two phenomena should be characterized by timescales of
the same order of magnitude; one involves the local dynamics of
baryons and non-baryonic dark matter, while the other involves
dark energy and the overall matter density (although see
\cite{kt} for a suggested explanation).
It is natural to wonder whether this is simply a numerical
coincidence, or the reflection of some new underlying theory
characterized by a single dimensionful parameter.  To date,
nobody has succeeded in inventing a theory which comes anything
close to explaining away both the dark matter and dark energy
in terms of modified gravitational dynamics.  Given the
manifold successes of the dark matter paradigm, from gravitational
lensing to structure formation to CMB anisotropy, is seems a good
bet to think that this numerical coincidence is simply an 
accident.  Of course, given the incredible importance of finding
a successful alternative theory, there seems to be little harm in
keeping an open mind.

It was mentioned above, and bears repeating, that modified-gravity
models do not hold any unique promise for solving the
coincidence problem.  At first glance we might hope that an
alternative to the conventional Friedmann equation might lead to
a naturally occurring acceleration at all times; but a moment's
reflection reveals that perpetual acceleration is not consistent
with the data, so we still require an explanation for why the
acceleration began recently.  In other words, the observations
seem to be indicating the importance of a fixed scale at which
the universe departs from ordinary matter domination; if we
are fortunate we will explain this scale either in terms of 
combinations of other scales in our particle-physics model or
as an outcome of dynamical processes, while if we are
unfortunate it will have to be a new input parameter to our
theory.  In either case, finding the origin of this new scale is
the task for theorists and experimenters in the near future.

\section{Discussion}

The discovery of dark energy has presented both observational
and theoretical cosmologists with a win-win scenario.  On the
observational side, we will either verify to high precision the
existence of a truly constant vacuum energy representing a new
fundamental constant of nature and a potentially crucial clue
to the reconciliation of gravity with quantum field theory, 
or we will detect variations
in the dark energy density which indicate either a new dynamical
component or an alteration of general relativity itself.  On
the theoretical side, we have been given invaluable insight into
one of the most perplexing issues in theoretical physics (the
cosmological constant problem), and we are now faced with a brand
new issue (the coincidence problem) whose resolution will 
necessarily involve exciting new theoretical developments.

Nobody who took arguments of naturalness and fine-tuning seriously
would have expected to discover a small but nonzero dark energy
density\footnote{This is a rhetorical exaggeration.
It has been pointed out to me, quite correctly,
that people who took naturalness arguments seriously might have
expected a nonzero cosmological constant if they also took the
anthropic principle seriously.}.  
We should not conclude from this that such arguments
have no value, but that we should always be prepared for
surprises.  One way of characterizing our current inventory of
the universe is to divide it into ordinary baryonic matter,
comprising $5\%$ of the energy density of the universe, and
a ``dark sector'' comprising the remaining $95\%$.  In this
classification, the role of the recent discoveries has been to
reveal that the dark sector includes at least two distinct
components, the dark matter and the dark energy.  Who is to say
that future experiments will not reveal further structure within
this sector, perhaps including interesting interactions between
components?  It is safe to say that the future of dark physics
looks very bright.

\section*{Acknowledgments}

This work was supported in part by the 
U.S. Dept.\ of Energy, the Alfred P. Sloan Foundation, and the David
and Lucile Packard Foundation.

\end{document}